\documentclass[aps,prc,twocolumn,superscriptaddress,showpacs,fleqn]{revtex4}
\usepackage[dvips]{graphicx}
\usepackage{amsmath,amssymb}

\begin{document}
\title{
Precession Mode on High-\mbox{\boldmath$K$} Configurations:
Non-Collective Axially-Symmetric Limit of Wobbling Motion
}

\author{Yoshifumi R. Shimizu}
\affiliation{Department of Physics, Graduate School of Sciences,
Kyushu University, Fukuoka 812-8581, Japan}

\author{Masayuki Matsuzaki}
\affiliation{Department of Physics, Fukuoka University of Education,
Munakata, Fukuoka 811-4192, Japan}

\author{Kenichi Matsuyanagi}
\affiliation{Department of Physics, Graduate School of Science,
Kyoto University, Kyoto 606-8502, Japan}

\maketitle

\leftline{\bf Abstract}
\vspace{2mm}
{\small
We have studied the precession mode,
the rotational excitation built on the high-$K$ isomeric state,
in comparison with the recently identified wobbling mode.
The random-phase-approximation (RPA) formalism, which has been developed
for the nuclear wobbling motion, is invoked and the precession phonon
is obtained by the non-collective axially-symmetric limit of the formalism.
The excitation energies and the electromagnetic properties
of the precession bands in $^{178}$W are calculated,
and it is found that the results of RPA calculations well correspond
to those of the rotor model; the correspondence can be
understood by an adiabatic approximation to the RPA phonon.
As a by-product, it is also found that
the problem of too small out-of-band $B(E2)$
in our previous RPA wobbling calculations can be solved by a suitable
choice of the triaxial deformation which corresponds to the one used
in the rotor model.
}

\vspace{6mm}
\leftline{\bf 1. Introduction}
\vspace{2mm}

In this talk, we present a recent progress of our study
on the wobbling and precession modes in nuclei.
These keywords, wobbling and precession, represents the motions
of classical tops.  They are quite interesting because they describe
three-dimensional (3-D) rotational motion,
and so related to a fundamental question:
How does a nucleus rotate as a 3-D quantum object?

We do not know how strictly these two are distinguished, but
in the following we use wobbling for motions of triaxial-body
and precession for those of axially-symmetric body.
In the lowest energy motion, the top rotates about one of the principal axes
with largest moment of inertia, but when excited the angular momentum vector
tilts from this axis in the body-fixed (intrinsic) frame.
Then, looking it from the laboratory frame, the body wobbles or precesses,
and that is why these names came from.  In the classical mechanics,
these two are similar; actually the difference is
that the trajectory of the angular momentum vector
is a pure circle in the case of precession, while it is an ellipse
in the case of wobbling.

However, the atomic nucleus is a quantum system and the situation is
dramatically changed: The collective rotation cannot occur about
the symmetry axis.  Thus, the quantum spectra corresponding
to the wobbling and precession motions are completely different.
In the case of wobbling, the spectra associated with an intrinsic
configuration are composed of multiple rotational bands;
the lowest (yrast) one represents the uniform rotation about
the main rotation axis, the first (one-phonon) excited band
represents a quantized motion of tilting angular momentum vector,
and so on (more excited band with more tilting angle).
These (${\mit\Delta}I=2$) rotational bands corresponds to
a collective rotation about the main rotation axis with largest
moment of inertia.  Since the excitation of phonons, or tilting
the angular momentum vector is another type of rotation about
the axis perpendicular to the main rotation axis,
these excitations form (${\mit\Delta}I=1$) vertical sequences.
In this way, the wobbling motions show a complicated band structure,
the horizontal rotational sequences and vertical phonon-like
excitations.

On the other hand, the angular momentum along the symmetry axis
(the main rotation axis) is generated by quasi-particle alignments
in the high-$K$ isomeric configurations,
and no collective rotation exists about it in the case of precession.
There are no horizontal sequences leaving only one vertical band
for each intrinsic configuration, which is nothing but a collective
rotation about the perpendicular axis to the high-$K$ angular momentum.

Recently, the wobbling phonon spectra have been identified
among the triaxial superdeformed (TSD) bands in Lu nuclei~\cite{Od01,Jen04},
up to two-phonon excitations.  By using the microscopic framework,
the random-phase approximation (RPA), we have studied the wobbling
phonon in the Hf-Lu region~\cite{MSM02,MSM04}.
The precession bands are rotational bands excited on
the prolate high-$K$ isomers and have been known for many years,
see e.g.~\cite{Cul99}.  We have recently investigated the properties of
the precession modes applying the same RPA formalism
in comparison with the wobbling modes~\cite{MS05,SMM05}.
The content of this talk is largely based on the results of~\cite{SMM05},
and a further development on the out-of-band $B(E2)$ transition
probability of wobbling phonon excitation (see \S3).

\vspace{6mm}
\leftline{\bf 2. Precession mode as a phonon}
\vspace{2mm}

The study of wobbling and precession modes has a long history
(see references quoted in~\cite{SMM05}).
Our recent works on the wobbling motions rely on the
microscopic framework developed in 1979 by Marshalek~\cite{Mar79}.
In almost the same time, a very important work for the precession mode
have been done here in Lund in 1981 by Andersson et al.~\cite{And81}
(c.f. also Refs.~\cite{Kura80,Kura82}).  In these works, was used
the microscopic RPA formalism, which is known to be suitable
to describe vibrational excitations. The reason why the RPA is
employed for such specific rotational motions as the wobbling
and precession modes can be easily understood in the case of
precession band, i.e. the high-$K$ rotational band.

The spectra of high-$K$ rotor is well-known:
\begin{equation}
  E_{{\rm high-}K}(I)=\frac{1}{2{\cal J}_\bot}[I(I+1)-K^2],
    \quad (I \ge K).
\label{eq:EhighK}
\end{equation}
Putting $n=I-K$ and assuming $K$ is large, it is easy to see that
this leads to an approximately harmonic spectrum,
\begin{equation}
  E_{{\rm high-}K}(n)=\omega_{\rm prec}
   \left[n+\frac{1}{2}+ \frac{n(n+1)}{K}\right],
\label{eq:EhoK}
\end{equation}
where $\omega_{\rm prec}\equiv K/{\cal J}_\bot$
is the precession phonon energy.
The remaining anharmonic term is of order $(1/K)$ and can be neglected
at the high-$K$ limit.  This harmonic picture is also valid for
the $E2$ transitions,
\begin{equation}
 B(E2) = (5/16\pi)\,Q_0^2\,\langle I_f K 2 0|I_i K \rangle^2.
\label{eq:BE2rot}
\end{equation}
By taking the high-$K$ asymptotic limit in the Clebsch-Gordan
coefficient,
$$
 \left\{ \begin{array}{l}
 B(E2; n+1\rightarrow n) \propto 3[(n+1)/K],
 \vspace{2mm} \\
 B(E2; n+2\rightarrow n) \propto (3/2)[(n+2)(n+1)/K^2],
 \end{array}\right.
$$
in which the one-phonon transition is order $(1/K)$ and
the direct transition to two-phonon state is $(1/K^2)$, so that
the direct two-phonon transitions are prohibited in the high-$K$ limit.

Now it is clear that the phonon picture is valid and the RPA formalism
can be used to describe the precession mode microscopically, which
has been done by Andersson et al.~\cite{And81}.  The key of their work
is to adopt the so-called symmetry-restoring separable type interaction:
The residual interaction used in the RPA is constructed in such a way
that the rotational symmetry broken by the deformed mean-field hamiltonian
is recovered.  This uniquely determines the complete form of the residual
interaction, and there is no adjustable parameter in the RPA calculation.
The dispersion equation is simply given as $\omega\,S(\omega)=0$,
where the $\omega=0$ solution is the rotational symmetry-restoring
mode, i.e. the Nambu-Goldstone (NG) mode, $J_{\pm}=J_y \pm iJ_z$.
Here and in the following
we assume the $x$-axis as the high-$K$ alignment axis.
The equation for non NG modes can be formally cast into the same form
as $\omega_{\rm prec}=K/{\cal J}_\bot$,
\begin{equation}
 \omega_{\rm prec}=\frac{K}{{\cal J}^{(\rm eff)}_\bot(\omega_{\rm prec})},
\label{eq:Wprec}
\end{equation}
but with microscopically defined moment of inertia,
$$
{\cal J}^{(\rm eff)}_\bot(\omega_{\rm prec})=\frac{1}{2}\sum_{\mu<\nu}
 \left\{
 \frac{|J_+(\mu\nu)|^2}
  {E_{\mu\nu}-\omega_{\rm prec}} +
 \frac{|J_-(\mu\nu)|^2}
  {E_{\mu\nu}+\omega_{\rm prec}} \right\}.
$$
However, the inertia ${\cal J}^{(\rm eff)}_\bot(\omega_{\rm prec})$ is
energy-dependent and actually Eq.~(\ref{eq:Wprec}) is a non-linear equation
to obtain the RPA eigen-energy.
Electromagnetic transitions can be calculated as the squared
amplitudes of the RPA phonon $X^\dagger_{\rm prec}$
with respect to appropriate transition operators $Q_{\lambda\mu}$:
\begin{equation}
 B(\lambda)_{{\mit\Delta}I}\approx\left|\langle K|
 [Q_{\lambda\mu={\mit\Delta}I}, X^\dagger_{\rm prec}]|K\rangle \right|^2
\,\,(K \gg 1).
\label{eq:BEL}
\end{equation}

Here it should be stressed that this RPA formalism of Andersson et al.
is obtained by taking the non-collective axially symmetric limit
($\gamma=60^\circ$ or $-120^\circ$ in the Lund convention) of the RPA wobbling
formalism of Marshalek~\cite{SMM05}.
In this way, the precession and wobbling
can be considered as similar kinds of collective excitation modes.

\begin{figure} [htbp]
\includegraphics[height=60mm,width=70mm,keepaspectratio,clip]{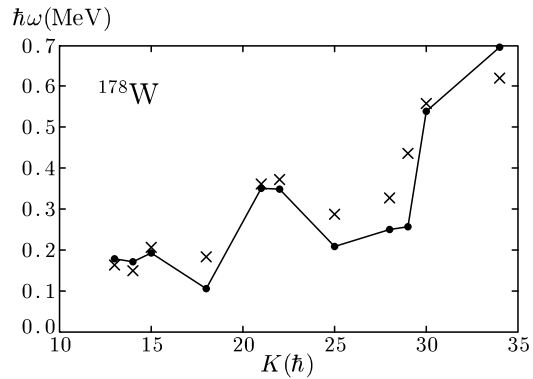}
\vspace{-3mm}
\caption{
 Excitation energies of the one-phonon precession modes
 excited on high-$K$ configurations.
 Calculated ones are denoted by filled circles connected by solid lines,
 and experimental ones by crosses. Taken from Ref.~\cite{SMM05}.
}
\label{fig:Ex}
\end{figure}

The results of RPA calculations for precession modes
in $^{178}$W~\cite{Cul99} are presented in Fig.~\ref{fig:Ex}.
In this nucleus there observed many high-$K$ isomers, on which
the rotational bands exist: Eleven isomers investigated are ranging
from four-quasiparticle states ($2\nu$-$2\pi$) to ten-quasiparticle states
($6\nu$-$4\pi$).  We have used the Nilsson potential as a mean field
with appropriate deformation parameters
and pairing gap parameters.
Except for four cases, $K^\pi=18^-,\,25^+,\,28^-,\,29^+$,
the one-phonon excitation energies are well reproduced.
As for these four isomers, the precession energies are too small;
namely, the calculated moments of inertia are too large.
We found that the overestimation of the calculated moments of inertia
is due to the proton contributions in these four configurations,
which include the $\pi[541]1/2^-$ Nilsson state
originated from the $\pi h_{9/2}$ high-$j$ decoupled orbit.
Apparently the contribution from this orbit is overestimated
in the present calculations,
see Ref.~\cite{SMM05} for more detailed discussions.

\begin{figure} [htbp]
\includegraphics[height=60mm,width=70mm,keepaspectratio,clip]{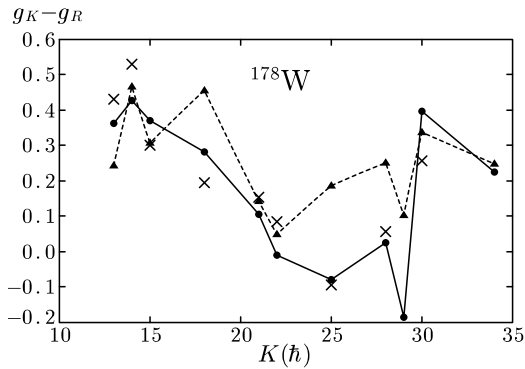}
\vspace{-3mm}
\caption{
 $(g_K-g_R)$-factors for high-$K$ precession bands.
 Those calculated by the RPA are denoted
 by filled circles connected by solid lines,
 while those by the mean-field approximation
 by filled triangles connected by dotted lines.
 Those extracted from the experimental data are
 shown by crosses.  Taken from Ref.~\cite{SMM05}.
}
\label{fig:gKgR}
\end{figure}

As for the electromagnetic transitions, one can directly compare
the calculated and measured transition probabilities, but here
we compare them in a different way.  As is well-known, the rotor model
gives simple formula for transition probabilities, e.g. Eq.~(\ref{eq:BE2rot})
for $B(E2)$.  Thus, by using the asymptotic forms
of the Clebsch-Gordan coefficients,
$B(E2)$ and $B(M1)$ are parameterized in the high-$K$ limit as,
\begin{eqnarray}
\label{eq:rotE2}
  B(E2:K+1\rightarrow K)_{\rm rot}
  &\approx& \frac{15}{16\pi}\frac{1}{K}Q_0^2, \\
  B(M1:K+1\rightarrow K)_{\rm rot}
  &\approx& \frac{3}{4\pi}\left(g_K-g_R\right)^2K.
\label{eq:rotM1}
\end{eqnarray}
On the other hand, they are calculated by Eq.~(\ref{eq:BEL})
in the RPA formalism.  By equating these two expressions,
we define $Q$ moment or $g_K-g_R$ factor calculated within the RPA.
Namely, we parameterize the results of RPA calculations in the same way
as the rotor model, and they are compared with experimentally extracted
ones in Fig.~\ref{fig:gKgR} for $B(M1)$.
In this figure we included the results of the simple mean-field approximation,
namely using $g_K$ calculated as
$\langle \mu_x \rangle/\langle J_x \rangle$ ($\mu_x$ is the $M1$ operator)
with each high-$K$ state, and a common value of $g_R$ calculated
in the same way but with the ground state.
As it is clearly seen in Fig.~\ref{fig:gKgR}, the RPA calculation
gives a much better description of $B(M1)$ values.
As for the $E2$ transitions, there is no experimental data available,
but by comparing the calculated RPA $Q$ moments
with the mean-field estimate of $Q$ moments it is found that
they well coincide except four configurations, which include
the $\pi h_{9/2}$ decoupled orbit
(the figure is not shown, see Ref.~\cite{SMM05}).

The results for the excitation energies and transition probabilities
indicate that the high-$K$ rotor model picture is realized if the model
parameters, moment of inertia, $Q$ moment and $g$-factors,
are calculated appropriately by means of the RPA.
This fact can be naturally understood by using
an adiabatic approximation studied in Refs.~\cite{Kura80,Kura82};
i.e. in this approximation,
$X_{\rm prec}^\dagger \approx
 (J_+ + K\,i{\mit\Theta}_+)_{\rm RPA}/\sqrt{2K}$,
while the NG mode is $ X_{\rm NG}^\dagger =J_-/\sqrt{2K}$.
By using this form it is easy to confirm, for example,
$(Q_0)_{\rm mean-field}\approx (Q_0)_{\rm RPA}$; the situation
is more complicated for the $g$-factors,
see Ref.~\cite{SMM05} for more detailed discussions.

\vspace{6mm}
\leftline{\bf 3. Out-of-band transition of wobbling phonon}
\vspace{2mm}

\begin{figure} [htbp]
\includegraphics[height=40mm,width=40mm,keepaspectratio,clip]{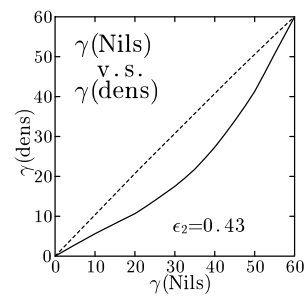}
\includegraphics[height=60mm,width=70mm,keepaspectratio,clip]{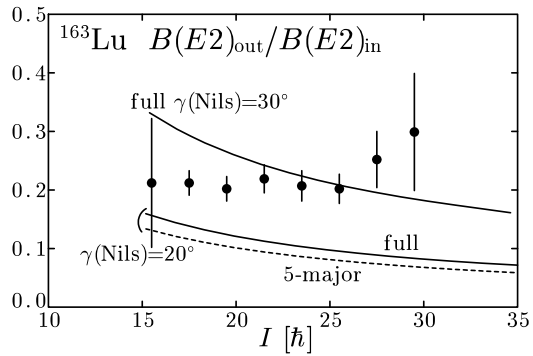}
\vspace{-3mm}
\caption{
 Upper panel: The relation between $\gamma({\rm Nils})$
 and $\gamma({\rm dens})$ for the TSD band in $^{163}$Lu.
 Lower panel: Comparison of the out-of-band to in-band $B(E2)$ ratios;
 symbols are the experimental data, three curves are corresponding
 results of RPA calculations, in which the dashed line is the result
 of Ref.~\cite{MSM02}, see text for detail.
}
\label{fig:BE2gamma}
\end{figure}

Now let us come back to the case of wobbling mode.
It has been shown that the RPA calculation well corresponds
to the rotor model picture in the case of precession,
especially for $B(E2)$.
It can be shown that a similar consideration
in term of the adiabatic approximation
can be also applied for the wobbling mode.
However, the $B(E2)$ value of our RPA wobbling calculations
in Lu nuclei were too small, by about 1/2 $-$ 1/3,
compared with the experimental data,
whose magnitudes are well reproduced by the rotor model.
This has been a serious problem for us,
see Fig.~\ref{fig:BE2gamma} and Refs.\cite{MSM02,MSM04}.
Considering the results of the precession modes, however,
it is difficult to understand the discrepancy between results of
the RPA calculation and the rotor model.  Therefore, we looked for
the reason why the RPA calculation gave such smaller values.

We have found that there are two reasons: One is related to
the model space of the RPA calculations, and the other is to
the definition of triaxiality parameter $\gamma$ in the Nilsson potential.
As for the model space, we used the 5-major shells but it was not
enough; this effect, however, is only about 20\% and not the major effect.
More important is the value of the $\gamma$ parameter.
It is believed that the $\gamma$ value is about $20^\circ$
in the TSD bands in the Lu region.
We have used $20^\circ$ for the $\gamma$
in the Nilsson potential, $\gamma({\rm Nils})=20^\circ$.
It is found, however, that the $\gamma$-value
defined by the density distribution, $\gamma({\rm dens)}$,
is only about $10^\circ$.  Here, $\gamma({\rm dens)}$ is defined by
the expectation values of the quadrupole operators (the Lund convention
for sign of $\gamma$),
$-\tan^{-1} \langle Q_{22}^{(+)}\rangle/\langle Q_{20}^{(+)}\rangle$, i.e.
\begin{equation}
 \gamma({\rm dens})=\tan^{-1}
 \frac{\sqrt{3}\,\displaystyle
 \langle\sum_{a=1}^A (y^2-x^2)_a\rangle }{\displaystyle
 \langle\sum_{a=1}^A (2z^2-x^2-y^2)_a\rangle }.
\label{eq:gamdens}
\end{equation}
Of course, the triaxiality of the rotor model
should be that of the density distribution.
By using an assumption for moments of inertia, ${\cal J}_y={\cal J}_z$,
in the triaxial rotor model,
the out-of-band to in-band $B(E2)$ ratio depends on
the $\gamma$ parameter like $B(E2)_{\rm out}/B(E2)_{\rm in}
\propto \tan^2{(\gamma+30^\circ)}$~\cite{Ham03}.
Even in general cases with ${\cal J}_y \ne {\cal J}_z$,
this $\gamma$-dependence approximately follows.
Then it is easy to see that the difference of this ratio
at $\gamma=10^\circ$ and $20^\circ$ is about factor two.
The $\gamma$ value corresponding to
$\gamma({\rm dens)}=20^\circ$ in the density distribution is
about $\gamma({\rm Nils)}\approx 30^\circ$ in the Nilsson potential
(see Fig.~\ref{fig:BE2gamma}).
The results of the RPA calculation using different $\gamma$-values
are shown in Fig.~\ref{fig:BE2gamma},
where the relation between $\gamma({\rm dens)}$ and $\gamma({\rm Nils)}$
is also depicted in the upper panel.
The dashed line with the model space of the 5-major shells is
the result of Ref.~\cite{MSM02}.   As is shown in the figure,
if one uses the full model space and $\gamma({\rm Nils)}=30^\circ$,
the $B(E2)$ ratio comes up to the correct magnitude
just like in the case of the rotor model~\cite{Gor04}.

The discrepancy between $\gamma({\rm Nils)}$ and $\gamma({\rm dens)}$
is just a matter of definition, see e.g. Appendix of Ref.~\cite{SM84}.
For the pure harmonic oscillator potential,
the definition of $\gamma({\rm Nils)}$ gives
\begin{equation}
 \gamma({\rm Nils}) =\tan^{-1} \frac{\sqrt{3}\,(\omega_y-\omega_x)}
 {2\omega_z -\omega_x-\omega_y}.
\label{eq:gamNils}
\end{equation}
The selfconsistency of the potential (Mottelson condition),
$\langle\sum_{a=1}^A (x_k^2)_a\rangle \propto 1/\omega_k^2$ ($k=x,y,z$),
is approximately satisfied in the Nilsson potential,
and then by Eq.~(\ref{eq:gamdens}),
\begin{equation}
 \gamma({\rm dens}) \approx \gamma({\rm self})
 = \tan^{-1} \frac{\sqrt{3}\,(1/\omega^2_y-1/\omega^2_x)}
 {2/\omega^2_z -1/\omega^2_x-1/\omega^2_y},
\label{eq:gamdensA}
\end{equation}
where $\gamma({\rm self})$ is the triaxiality based on the shape
(ellipsoid) of an equi-potential surface
for the anisotropic harmonic oscillator.
In this way the relation between
$\gamma({\rm Nils)}$ and $\gamma({\rm dens)}$ in the upper panel
of Fig.~\ref{fig:BE2gamma} can be naturally understood.

\vspace{6mm}
\leftline{\bf 4. Summary}
\vspace{2mm}

The RPA precession formalism by Andersson et al. can be obtained
by a non-collective axially-symmetric limit of the RPA wobbling
formalism by Marshalek.  By using this formalism, it is shown that
the RPA calculation gives fairly good agreements
for both the excitation energies
and $B(M1)$ transitions for the precession phonons
on high-$K$ isomers in $^{178}$W.
This result can be interpreted by an adiabatic approximation,
and shows a good correspondence between the RPA calculation
and the rotor model;
especially $B(E2)$ or $Q$ moments, and $B(M1)$ or $g$-factors~\cite{SMM05}.

There is an important feed back to the calculation
of recently observed nuclear wobbling motions.
The problem of small $B(E2)$ ratios
in our previous RPA calculations can be solved if one uses a proper value
of the triaxiality parameter $\gamma({\rm dens}) \approx 20^\circ$
in the density distribution.  This does not completely solve the problem,
however, because the Nilsson-Strutinsky calculation gives minima
at $\gamma({\rm Nils}) \approx 20^\circ$ in the Nilsson potential,
which corresponds to $\gamma({\rm dens}) \approx 10^\circ$.
Thus, it raises a more fundamental question;
why the Nilsson-Strutinsky calculation does not provide
enough triaxiality, which is required for explaining the measured
$B(E2)$ ratio of the wobbling phonon band.
Finally, it is shown that the RPA can describe the 3-D rotational
motion (in the small amplitude approximation) at least as the same level
as the rotor model.  In other words, the rotor model may be justified
by the microscopic RPA calculations.

\vspace{6mm}
\leftline{\bf Acknowledgments}
\vspace{2mm}

This work was supported by the Grant-in-Aid for Scientific Research
(No. 16540249) from the Japan Society for the Promotion of Science.

\end{document}